\documentclass[aps,prl,reprint]{revtex4-1}
\usepackage{amsmath,amsfonts,amssymb}
\usepackage{wrapfig}
\usepackage{graphicx}
\usepackage{bbm}
\usepackage{graphics}
\usepackage{float}
\usepackage{color}
\usepackage{braket}
\usepackage[hidelinks]{hyperref}

\begin{document}
\pagestyle{plain}

\title{Quadrature squeezed photons from a two-level system}

\author{Carsten H. H. Schulte}
\thanks{These authors contributed equally.}
\affiliation{Cavendish Laboratory, University of Cambridge, JJ Thomson Avenue, Cambridge CB3 0HE, United Kingdom}
\author{Jack Hansom}
\thanks{These authors contributed equally.}
\affiliation{Cavendish Laboratory, University of Cambridge, JJ Thomson Avenue, Cambridge CB3 0HE, United Kingdom}
\author{Alex E. Jones}
\affiliation{Cavendish Laboratory, University of Cambridge, JJ Thomson Avenue, Cambridge CB3 0HE, United Kingdom}
\author{Clemens Matthiesen}
\affiliation{Cavendish Laboratory, University of Cambridge, JJ Thomson Avenue, Cambridge CB3 0HE, United Kingdom}
\author{Claire Le Gall}
\affiliation{Cavendish Laboratory, University of Cambridge, JJ Thomson Avenue, Cambridge CB3 0HE, United Kingdom}
\author{Mete Atat\"ure}
\affiliation{Cavendish Laboratory, University of Cambridge, JJ Thomson Avenue, Cambridge CB3 0HE, United Kingdom}






\date{\today}

\maketitle

\textbf{Resonance fluorescence arises from the interaction of an optical field with a two-level system and has played a fundamental role in the development of quantum optics and its applications. Despite its conceptual simplicity it entails a wide range of intriguing phenomena, such as the Mollow-triplet emission spectrum\cite{Schuda1974}{}, photon antibunching\cite{Kimble1977} and coherent photon emission\cite{Hoffges1997}{}. One fundamental aspect of resonance fluorescence, reduced quantum fluctuations in the single photon stream from an atom in free space, was predicted more than 30 years ago\cite{Walls1981}{}. However, the requirement to operate in the weak excitation regime, together with the combination of modest oscillator strength of atoms and low collection efficiencies, has continued to cast stringent experimental conditions for the observation of squeezing with atoms. Attempts to circumvent these issues had to sacrifice antibunching due to either stimulated forward scattering from atomic ensembles \cite{Heidmann1985,Lu1998} or multi-photon transitions inside optical cavities \cite{Raizen1987,Ourjoumtsev2011}{}. Here, we use an artificial atom with a large optical dipole enabling 100-fold improvement of the photon detection rate over the natural atom counterpart\cite{Gerber2009} and reach the necessary conditions for the observation of quadrature squeezing in single resonance-fluorescence photons. Implementing phase-dependent homodyne intensity-correlation detection\cite{Ou1987,Vogel1991,Vogel1995,Gerber2009}{}, we demonstrate that the electric field quadrature variance of resonance fluorescence is 3\% below the fundamental limit set by vacuum fluctuations, while the photon statistics remain antibunched. The presence of squeezing and antibunching simultaneously is a fully nonclassical outcome of the wave-particle duality of photons.}

\noindent The minimum fluctuations in any quantum measurement of canonically conjugate variables such as position and momentum are bound by the Heisenberg uncertainty principle. Although this principle cannot be violated, the fluctuations of a single variable can be reduced below this minimum value at the expense of enhancing the fluctuations of the conjugate variable. The most widely explored realization of this nonclassical phenomenon is squeezed light \cite{Yuen1976} where the quadrature operators $\hat{X}_1$ and $\hat{X}_2$ of the electric field are the canonically conjugate operators. Relying inherently on the quadratic dependence on the bosonic creation and annihilation operators in the Hamiltonian, squeezed light is typically generated using intense lasers and macroscopic nonlinear-optical media \cite{Teich1989}{}. This form of squeezed light has multiple applications in the field of quantum optics\cite{Walls1983}{}, one prominent example being interferometry with reduced quantum noise\cite{Caves1982,Goda2008}{}.\\
In 1981, Walls and Zoller predicted that the quadratic form of the Hamiltonian is not a requirement and that quadrature squeezing can also be obtained via a radically different approach: the interaction of a two-level system with a resonant light field, as described by the Jaynes-Cummings Hamiltonian \cite{Walls1981}{}. The fluctuations in one quadrature, quantified by their variance, can be reduced up to a theoretical maximum of 12.5\% lower than vacuum fluctuations, while the intensity statistics remain antibunched. Unlike its nonlinear-optics counterpart this unique form of squeezed light stems from a buildup of atomic coherence which, once mapped onto the emitted field, results in the creation of coherences between the $n=0$ and $n=1$ Fock states in the weak excitation regime. Higher number states are excluded by photon antibunching or equivalently, by the fermionic nature of atomic operators. The simultaneous presence of antibunching and squeezing is an intriguing yet counter-intuitive effect, since single photons do not have a well-defined phase. It is the coherence with the zero-photon, i.e. vacuum, component that allows for phase-dependent effects such as the squeezing discussed here.\\
The two-level system we use in this work is a voltage-controlled semiconductor quantum dot (QD)\cite{Warburton2000} positioned under a solid immersion lens for enhanced photon collection efficiency \cite{Zwiller2002}{}. Typically allowing photon detection rates exceeding well above a million photons per second, these artificial atoms obviate the immediate need for cavity coupling and consequently allow for the experimental realisation of an isolated, weakly excited two-level system treated in ref. \cite{Walls1981}. With large oscillator strength, high internal quantum efficiency and negligible decoherence, semiconductor QDs enabled recent observations of key phenomena in quantum optics, such as antibunching \cite{Michler2000,Kim1999}{}, formation of dressed states \cite{Xu2007,Vamivakas2009,Flagg2009}{}, generation of entangled photon pairs \cite{Akopian2006,Young2006,Muller2014} and, particularly relevant for this work, coherent single photon generation via weak excitation\cite{Matthiesen2012,Matthiesen2013}{}. The strong transition dipole moment of the QD compared to single atoms is also the key enabler in the detection of squeezing in resonance fluorescence, since signal to noise ratio is fundamentally tied to the photon detection rate.\\
%
\begin{figure*}[ht]
\centering
\includegraphics[width=2\columnwidth]{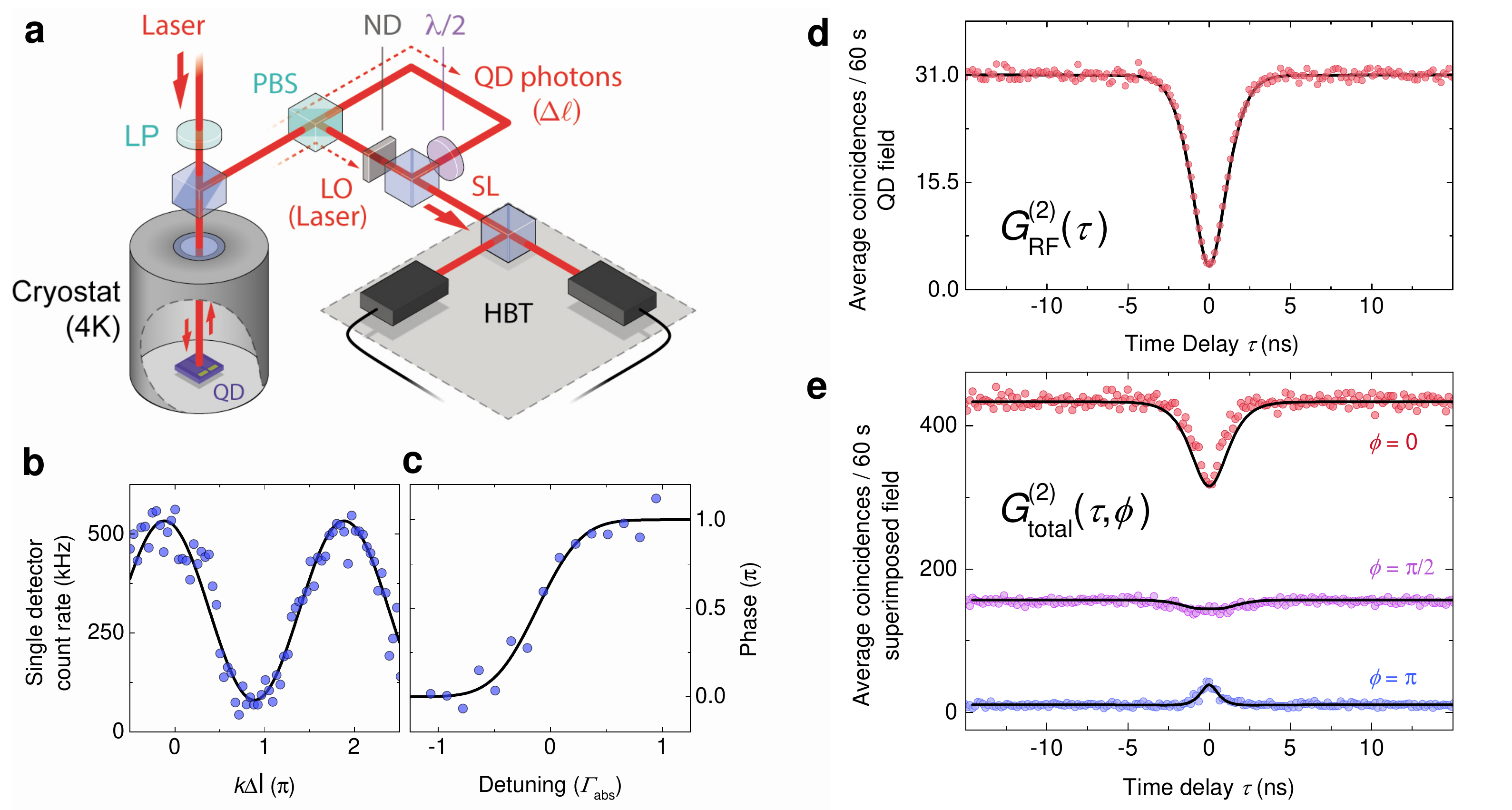}
\caption{{\bf Homodyne intensity-correlation detection.} \textbf{a,} Schematic illustration of the homodyne intensity-correlation setup. LP: linear polariser, PBS: polarising beam splitter, ND: neutral density filter, $\lambda/2$: half-wave plate. \textbf{b,} Intensity of the SL field on a single detector as a function of the interferometer phase at an excitation power of $s=0.1$. At this excitation power, each detector of the HBT setup records $1.6\cdot10^5$ events/s from resonance fluorescence contribution alone. \textbf{c,} Dipole phase offset in interference pattern as a function of detuning between laser and QD frequency. \textbf{d, e,} Intensity autocorrelation measurement with LO path blocked (\textbf{d}) and unblocked (\textbf{e}). In the blocked case, the $G_{\mathrm{RF}}^{(2)}(\tau)$ measurement gives the expected antibunching, evidence of a single two-level system, regardless of $\phi$. The ordinate in panels \textbf{d} and \textbf{e} shows the coincidences in units of count rates for comparison. For unblocked LO (panel \textbf{e}), the interference between LO and QD fields leads to phase-dependent correlations, some of which contain the quadrature variance of the QD field.}
\end{figure*}
%
To generate resonance fluorescence, we excite the $\pi^+$-polarised neutral exciton transition of a single QD resonantly at 970 nm using a frequency-stabilised tunable laser (Fig. 1a). Resonance fluorescence and the reflected laser are separated by a polarising beam splitter and, after imparting a relative phase through a path length difference $\Delta \ell$, recombined via a non-polarising beam splitter. One of the outputs of this beam splitter contains the superimposed light (SL) field\\
\begin{align}
\label{SLfield}
\hat{E}^{(+)}_{\mathrm{SL}}(t)=\hat{E}^{(+)}_{\mathrm{RF}}(t)+e^{i\phi}\cdot \hat{E}^{(+)}_{\mathrm{LO}}(t),
\end{align}
where the subscript LO (RF) indicates local oscillator (resonance fluorescence) and the relative amplitude and phase $\phi$ of the LO and RF fields can be tuned independently in the experimental scheme illustrated in Fig. 1a. Reflection and transmission coefficients and all other relative phases due to the optical setup are included in the field amplitudes and the phase $\phi$. Figure 1b shows the intensity measured on a single detector as a function of the interferometer-induced phase $k\Delta \ell$, where $k$ is the wavenumber of the fields. The phase due to the dipolar response of the transition, which is determined by the relative detuning between the excitation laser and the transition frequency, is also included in $\phi$; Fig. 1c displays the measurement of the detuning dependence of this additional phase.\\
The amplitude $\left<\hat{E}(\phi)\right>$ of a light field, where $\phi$ is the relative phase with respect to a coherent reference field, can be represented in the phase space of the conjugate variables, $\hat{X}_1$ and $\hat{X}_2$, via $\hat{E}(\phi)\propto\hat{X}(\phi)=(\hat{X}_1 \cos{\phi}+\hat{X}_2 \sin{\phi})$. These quadratures are the analogs of the dimensionless position and momentum operators and their variances, quantifying the quantum fluctuations of the electric field, are subject to a similar uncertainty relation:
\begin{equation}
\label{HeisenbergQuadratures}
\Delta\hat{X}_1^2 \Delta\hat{X}_2^2 \geq 1/16.
\end{equation}
%
\begin{figure*}[ht]
\centering
\includegraphics[width=2\columnwidth]{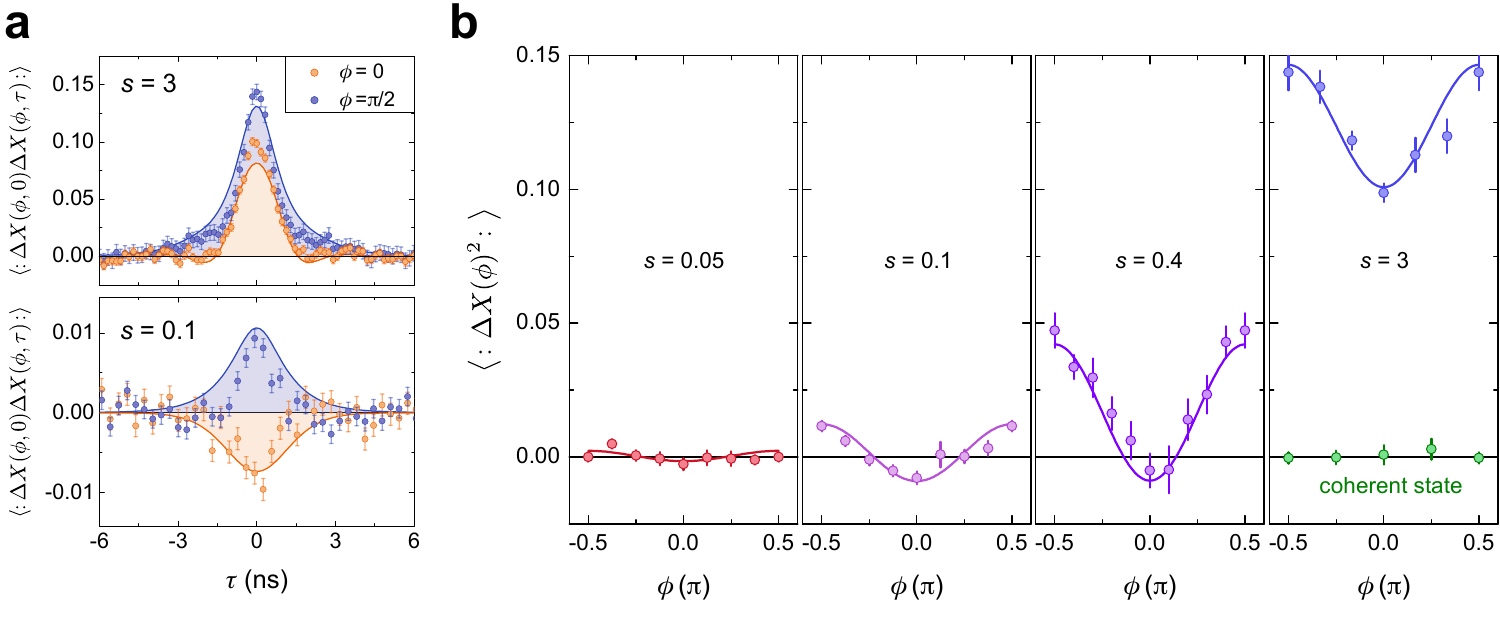}
\caption{{\bf Phase-dependent quadrature variance of resonance fluorescence.} \textbf{a,} In-phase (orange) and out-of-phase (blue) normally ordered autocorrelations of the electric field quadrature fluctuations $\left<\colon(\Delta \hat{X}(\phi,0)\Delta\hat{X}(\phi,\tau))\colon \right>$ for high and low power excitation conditions are displayed in top and bottom panels, respectively. Negative values in panel \textbf{a} (bottom) verify squeezing of the in-phase electric field variance. \textbf{b,} Full phase dependence of the quadrature variances (zero-time delay of the autocorrelations), for different excitation powers. A measurement of coherent laser quadratures provides a reference for the vacuum limit of zero (green circles, rightmost panel). Solid curves in all panels are theoretical simulations using a two-level Master equation for the corresponding experimental conditions.}
\end{figure*}
%
To demonstrate a squeezed quadrature, i.e. $\Delta\hat{X}(\phi)^2<1/4$, we implement the experimental setup proposed in Ref. \onlinecite{Vogel1995}{}, which provides a direct and convenient link between time-correlated two-photon detection and the variances of field quadratures. To detect the variances of the RF field quadratures $\hat{X}_{1,2}$, we perform an intensity autocorrelation on the SL field $\hat{E}_{\mathrm{SL}}$ using a Hanbury Brown and Twiss correlation setup (HBT). The unnormalised second-order correlation function of the SL field,
\begin{equation}\label{Gtotal}
G^{(2)}_{\textrm{total}}(t,t+\tau)=\left<\hat{E}^{(-)}_{\mathrm{SL}}(t)\hat{E}^{(-)}_{\mathrm{SL}}(t+\tau)\hat{E}^{(+)}_{\mathrm{SL}}(t+\tau)\hat{E}^{(+)}_{\mathrm{SL}}(t)\right>.
\end{equation}
produces the well-known antibunched second-order correlation function of the resonance fluorescence field, $G^{(2)}_{\mathrm{RF}}$, in the absence of a local oscillator, showing the single photon nature of resonance fluorescence\cite{Kimble1977}{}. The solid red circles in Fig. 1d display this behaviour for an excitation power of $s=P/P_{\mathrm{sat}}=0.1$, where the saturation power $P_{\mathrm{sat}}$ yields half of the maximum attainable resonance fluorescence intensity. The black curve is the theoretical prediction obtained with a two-level Master equation and includes the detector response function, as well as sub-linewidth spectral wandering of the QD transition frequency \cite{Kuhlmann2013,Stanley2014}{}. In the presence of the LO field, $G^{(2)}_{\textrm{total}}$ displays a strong dependence on phase $\phi$, as shown in Fig. 1e for $\phi=0,\;\pi/2$ and $\pi$. While the coincidence rate at long time delays changes with $\phi$ by more than an order of magnitude, the correlation behaviour at short time delays evolves from a dip to a peak as a function of $\phi$.\\
The total correlation function $G^{(2)}_{\textrm{total}}$ contains five terms with $\left|E_\mathrm{LO}\right|^n$ for $n$ ranging from 0 to 4. We can separate their contributions via their unique dependence on the time delay ($\tau$) and relative phase ($\phi$), as well as via direct measurement of the zeroth-order contribution (Fig. 1d) (see SI). The second-order contribution is directly proportional to the normally ordered autocorrelation of $\Delta X(\phi,t)$:
\begin{equation}\label{eq:fluctuationautocorrelation}
\Delta G_2^{(2)}(\tau)\propto\left<\colon\Delta \hat{X}(\phi,0)\Delta \hat{X}(\phi,\tau)\colon\right>.
\end{equation}
The zero-delay value of Eq. \ref{eq:fluctuationautocorrelation} hence yields the normally ordered variance $\left<\colon\left(\Delta \hat{X}(\phi)\right)^2\colon\right>$ of the electric field quadrature \cite{Vogel1995}{}. This variance is zero for vacuum and for coherent fields and the existence of quadrature squeezing is manifested in a negative-valued normally ordered variance:
\begin{equation}\label{eq:squeezingcondition}
\left<\colon\left(\Delta \hat{X}(\phi)\right)^2\colon\right> < 0.
\end{equation}
Figure 2a presents the autocorrelation of the in-phase ($\phi=0$) and in-quadrature ($\phi=\pi/2$) fluctuations for high excitation power ($s=3$, upper panel). The normally ordered variance ($\tau=0$) of resonance fluorescence in the high power regime is positive valued regardless of the phase. This indicates that the quantum fluctuations are enhanced above the vacuum level, as expected. In stark contrast, the low power regime ($s=0.1$), shown in the lower panel, yields negative values for the normally ordered variance for $\phi=0$. This reduction of quantum fluctuations below the vacuum limit is the verification of quadrature squeezing in this measurement. As dictated by the Heisenberg uncertainty relations, this squeezing is accompanied by increased fluctuations, i.e. antisqueezing, in the other quadrature. Both features decay on a timescale of the order of the excited state lifetime \cite{Loudon1984}{}.\\
Figure 2b shows how the normally ordered variance evolves as a function of $\phi$, for excitation powers ranging from $s=0.05$ (leftmost panel) to $s=3$ (rightmost panel). A measurement with a weak laser of similar intensity is displayed in the rightmost plot as green circles. As expected for a coherent state, this measurement yields a normally ordered variance of zero, independent of $\phi$. The squeezing vanishes \cite{Walls1981} for $s\geq1$, yielding larger fluctuations than vacuum for any $\phi$. While we observe the phase dependence of the quadrature variance at all excitation powers, the window of opportunity for measuring negative values is restricted to a very small $\phi$-range in the low-excitation regime, i.e. $s<1$, highlighting the challenges for the observation of squeezing in resonance fluorescence since its prediction.
%
\begin{figure}[ht!]
\centering
\includegraphics[width=\columnwidth]{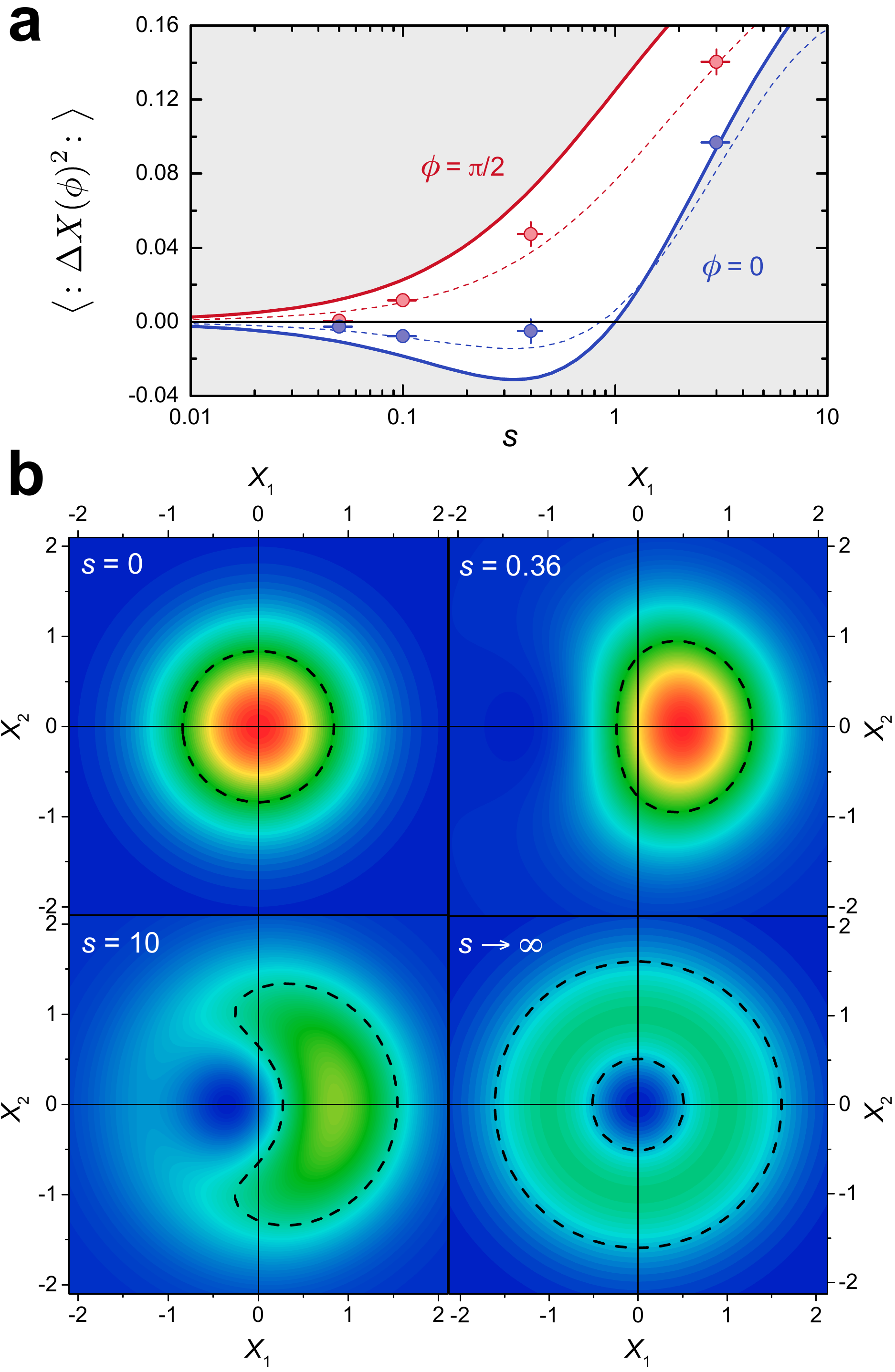}
\caption{{\bf Excitation-power dependence of quadrature squeezing.} \textbf{a,} The measured (symbols) and the theoretical (solid curves) normally ordered quadrature variances as a function of the excitation power $s$ is shown for $\phi=0$ and $\phi=\pi/2$. The dashed curves illustrate the effect of phase noise and finite timing resolution in our experiment on an otherwise ideal two-level system (see SI). \textbf{b,} Wigner functions for different excitation powers. The transition from vacuum state ($s=0$) to a mixture of vacuum and single photon Fock state ($s\rightarrow\infty$) displays non-symmetric forms at intermediate excitation regimes ($s=0.36$ and $s=10$). The panel for $s=0.36$ displays clearly that the spread of the Wigner function along $X_{1}$ is less than that of vacuum, a manifestation of quadrature squeezing. The dashed lines depict the contour at 50\% of the maximum value at each power.}
\end{figure}

\noindent Figure 3a summarises the power dependence of the normally ordered quadrature variance extrema. The solid blue (red) curve represents the theoretically predicted behaviour of the in-phase (in-quadrature) field variance for an ideal two-level system. The maximum possible squeezing is limited to 12.5\% (0.58 dB) below vacuum fluctuations at $s=0.36$. The dashed curves depict how the ideal two-level system behaviour is modified due to the combined effects of finite timing resolution of our detection system and phase uncertainty of our interferometer (see Methods). All variance extrema we measure are commensurate with these predictions confirming that deviation from the solid curve is predominantly of technical nature. The maximum degree of squeezing we measure is $3.1\pm1$\% (0.14 dB) below vacuum noise at an excitation power of $s=0.1$. This value corresponds to 40\% of the theoretically obtainable limit set by the blue solid curve at this excitation power.\\
The transformation of the state of light with excitation power is best visualised by the calculated Wigner functions presented in Fig. 3b. The spread of these phase-space distributions along a given polar angle, $\phi$, is indicative of the variance of the corresponding field quadrature $\hat{X}(\phi)=\hat{X}_{1}\cos{(\phi)}+\hat{X}_{2}\sin{(\phi)}$. The Wigner function for the vacuum state (top left) shows a symmetric form with no phase dependence. At intermediate powers (top right and bottom left), the symmetry breaks down and a $\phi$-dependence arises in the spread of the Wigner function, linked to the generation of atomic coherence (see SI). This phase dependence, in combination with the antibunched nature of resonance fluorescence, leads to a reduced variance of the electric field for a phase angle of $\phi = 0$ (c.f. Wigner function at $s=0.36$). In the high power regime ($s\rightarrow\infty$), the field becomes a statistical mixture of $n=0$ and $n=1$ Fock states and the steady-state phase dependence is lost completely.\\
We have shown that resonance fluorescence from a two-level system can comprise a single photon stream with below-vacuum quantum fluctuations of the field. While this appears counterintuitive owing to the impossibility of associating a well-defined phase to single photons, the probabilistic nature of coherent photon scattering in the weak excitation regime allows the emitted photons to be in a coherent superposition of Fock states $\left|0\right>$ and $\left|1\right>$. The emergence of phase correlations in this regime endows resonance fluorescence with the coexistence of photon antibunching and quadrature squeezing. Our simultaneous observation of these two phenomena can therefore be interpreted as a quantum mechanical manifestation of the complementary particle and wave natures of light, respectively, with no classical analogs.

\begin{acknowledgments}
We gratefully acknowledge financial support by the University of Cambridge, the European Research Council ERC Consolidator Grant Agreement No. 617985 and the EU-FP7 Marie Curie Initial Training Network S3NANO. C.M. gratefully acknowledges Clare College Cambridge for financial support through a Junior Research Fellowship. We thank E. Clarke, M. Hugues and the EPSRC National Centre for III-V Technologies for the wafer and C. Baune, R. Moghadas Nia, A. Ourjoumtsev, W. Vogel and H. J. Carmichael for fruitful discussions.
\end{acknowledgments}
Correspondence and requests for materials should be addressed to M.A. (ma424@cam.ac.uk).

\section{Methods}

\noindent \textbf{Interferometer:}  A frequency- and power-stabilised single-mode laser is used to resonantly excite the neutral exciton transition of the QD. The emitted photons are collected in a confocal dark-field microscope, where the laser is separated from the QD emission by means of two crossed polarisers. The second polariser is implemented as a polarising beam splitter (PBS), which enables the use of the attenuated laser field as local oscillator (LO). The light field in the QD arm of the interferometer consists typically of $<1$\% laser photons and $>99$\% RF photons, we therefore neglect the laser background in the QD-photon mode. Likewise, any QD photons in the LO-output can be neglected because the excitation laser intensity before attenuation is several orders of magnitude larger than the RF intensity. In the fringe measurement in Fig. 1a the spatial path difference $\Delta \ell\approx 11$ cm is kept constant while laser and QD frequency are tuned continuously to change the interferometer phase. This form of phase control is enabled by the tuneability of the QD resonance via the DC Stark effect \cite{Warburton2002} and increases the long-term stability of the interferometer, which contains no moving parts. The visibility of the interferometer for high power laser light amounts to near unity but is reduced in Fig. 1b to 73.8\% due to incoherent photon emission as well as an inadvertent mismatch of RF and LO intensities. Additionally, the visibility is reduced for low count rates \cite{Gerber2009}{}, which makes the use of a bright single photon source and high photon collection efficiencies crucial for our experiments. The collection efficiency of our optical setup is $1$\%. This value is calculated from the obtained count rates and the $0.58-$ns excited state lifetime of the QD used in this work.

\noindent \textbf{Postselection:} The intensities of RF and LO fields are kept equal in all of our measurements. To ensure the absence of laser photons in the QD mode, the laser background is measured once a minute for 2 s. To this end, the LO path is blocked and the QD, which is embedded in a Schottky diode structure, is tuned off resonance via the quantum confined Stark effect (QCSE) \cite{Warburton2002}{}. Furthermore, the intensity of the QD emission is monitored continuously during the measurements to detect spectral wandering of the QD transition \cite{Kuhlmann2013,Matthiesen2014}{}. This is done by filtering out the phonon sideband (PSB) and detecting it on a third single photon detector \cite{Hansom2014}{}. The measured correlation histograms ($G^{(2)}_\mathrm{total}(\tau)$) are saved once a minute. We perform a postselection of histograms with a threshold on the mean PSB count rate and another threshold on the measured laser leakage in the QD arm. In the experiments shown in Fig. 2, the laser is kept on resonance but the relative phase is not actively controlled in the interferometer. Instead, phase-dependent measurements are performed by using the individual detector intensities as a measure of the interferometer phase and relying on the wandering of the phase due to temperature drifts on timescales of typically $\gtrsim30$ min/$\pi$. In order to bin the data we perform a reference measurement of the interference fringes by scanning the laser frequency while keeping the QD on resonance using the QCSE. An example measurement of the interference fringes obtained in this way can be seen in Fig. 1b. In order to have equal sized phase bins, we use intensity bins of varying size proportional to the derivative of a $\cos^2(\phi/2)$ function. We note that this phase binning is sign-invariant, i.e. it cannot distinguish between positive or negative phases and therefore bins data into values between 0 and $+\pi$. This does not impact our measurement as all correlation functions are symmetric in phase around 0. The data points shown at negative phases in Fig. 2b are measured between $\phi=0.5\pi$ and $\phi=\pi$, and have been shifted by $-\pi$.

\noindent \textbf{Reduction of measured degree of squeezing:} Although the conditioning nature of the measurement should render it robust against low detection efficiencies, this is not true in practice, and several effects reduce the measured degree of squeezing compared to the theoretical limit for resonance fluorescence. Low photon numbers in the interferometer reduce the fringe visibility, but this only affects the signal-to-noise ratio in our measurements. However, low count rates also lead to higher shot noise which can increase the error in the phase binning protocol. This in turn leads to a decrease in the detected degree of squeezing by introducing mixing between quadratures \cite{Zhang2003}{}. Other sources of phase noise include spectral wandering of the QD transition leading to fluctuations in the dipole phase, and any interferometric instability on timescales shorter than the histogram saving time. Finite timing resolution of the correlation setup also leads to a decreased visibility of the features at 0 time delay and further reduce our measured value for squeezing. We have independently measured the timing resolution of the HBT setup with a mode-locked pulsed laser source ($< 3$ ps pulse width) for different mean count rates. The extent of phase-noise from different sources is harder to quantify and is used as a fitting parameter in the theoretical curves in Fig. 2.

\bibliography{squeezing-noURLs}
\bibliographystyle{naturemag}

\end{document}